\def\ra{\rangle}
\def\la{\langle}
\def\be{\begin{equation}}
\def\ee{\end{equation}}
\def\ba{\begin{array}}
\def\ea{\end{array}}

\documentclass[prl,showpacs,twocolumn,amsmath]{revtex4}
\usepackage{epsfig,graphicx}

\begin{document}

\title{Genuine multipartite entanglement detection and
lower bound of multipartite concurrence}
\author{Ming Li$^{1,2}$}
\author{Shao-Ming Fei$^{2,3}$}
\author{Xianqing Li-Jost$^{2}$}
\author{Heng Fan$^{4,5}$}
\affiliation{$^1$College of the Science, China University of
Petroleum, 266580 Qingdao, China\\
$^2$ Max-Planck-Institute for Mathematics in the Sciences, 04103
Leipzig, Germany\\
$^3$ School of Mathematical Sciences, Capital Normal University, Beijing 100048, China\\
$^4$ Beijing National Laboratory for Condensed Matter Physics,
Institute of Physics, Chinese Academy of Sciences, Beijing 100190,
China\\
$^5$ Collaborative Innovative Center of Quantum Matter, Beijing
100190, China}

\begin{abstract}
The problems of genuine multipartite entanglement detection and classification are challenging.
We show that a multipartite quantum state
is genuine multipartite entangled if the multipartite concurrence is larger than
certain quantities given by the number and the dimension of the subsystems. This result also
provides a classification of various genuine multipartite entanglement.
Then, we present a lower bound of
the multipartite concurrence in terms of bipartite concurrences. While various
operational approaches are available for providing lower bounds of bipartite concurrences,
our results give an effective operational way to detect and classify
the genuine multipartite entanglement.
As applications, the genuine multipartite entanglement of tripartite systems is analyzed in detail.

\end{abstract}

\smallskip

 \maketitle

Quantum entanglement, as the remarkable nonlocal feature of quantum
mechanics, is recognized as a valuable resource in the rapidly
expanding field of quantum information science, with various
applications \cite{nielsen, di} in such as quantum computation,
quantum teleportation, dense coding, quantum cryptographic schemes,
quantum radar, entanglement swapping and remote states preparation.
A bipartite quantum state without entanglement is called separable.
A multipartite quantum state that is not separable with respect to
any bi-partitions is said to be genuinely multipartite
entangled \cite{dct,huber,vicente3}. Genuine multipartite
entanglement is an important type of entanglement, which offers
significant advantages in quantum tasks comparing with bipartite
entanglement \cite{mule1}. In particular, it is the basic ingredient
in measurement-based quantum computation \cite{mule2}, and is
beneficial in various quantum communication protocols \cite{mule3},
including secret sharing \cite{mule4} (cf. \cite{mule5}), among
multi-parties. Despite its importance, characterization and
detection of this kind of resource turns out to be quite difficult.
Recently some methods such as linear and nonlinear entanglement
witnesses \cite{12, huber, vicente3, huber1, wu, sperling, 14, 15,
claude, horo}, generalized concurrence for multipartite genuine
entanglement \cite{ma1, ma2,gaot1,gaot2}, and Bell-like inequalities
\cite{bellgme} have been proposed. Nevertheless, the problem remains
far from being satisfactorily solved.

Quantifying entanglement is also a basic and long standing problem
in quantum information theory \cite{wootters,multicon1,multicon2,chenk}. Estimation of any
quantum entanglement measures can be used to judge the separability of a given state.
From the norms of the correlation tensors in the generalized Bloch
representation of a quantum state, separable conditions for both bi- and multi-partite quantum states
are presented in \cite{vicente1,vicente2,hassan,ming}; a multipartite
entanglement measure for N-qubit and N-qudit pure states is given in
\cite{hassan1,hassan2};  a general framework for detecting genuine multipartite entanglement
and non full separability in multipartite quantum systems of
arbitrary dimensions has been introduced in \cite{vicente3}.
In \cite{mingbell} it has been shown that the norms of the correlation tensors has
a close relationship to the maximal violation of a kind of multi Bell inequalities.

In this Letter, we investigate the genuine multipartite entanglement
in terms of the norms of the correlation tensors and multipartite
concurrence. We show that if the multipartite concurrence is larger
than a constant given by the number and dimension of the subsystems,
the state must be genuine multipartite entangled. To implement the
criteria, we investigate the relationship between the bipartite
concurrence and the multipartite concurrence. An effective lower
bound of multipartite concurrence is derived to detect genuine
multipartite entanglement.

Let ${\mathcal {H}}_{i}$, $i=1,2,...,N$, denote $d$-dimensional Hilbert spaces.
The concurrence of an $N$-partite quantum pure state $|\psi\ra\in {\mathcal
{H}}_{1}\otimes{\mathcal {H}}_{2}\otimes\cdots\otimes{\mathcal
{H}}_{N}$ is defined by \cite{multicon1,multicon2},
\begin{eqnarray}\label{xxx}
C_{N}(|\psi\ra\la\psi|)=2^{1-\frac{N}{2}}\sqrt{(2^{N}-2)-\sum_{\alpha}Tr\{\rho_{\alpha}^{2}\}},
\end{eqnarray}
where $\alpha$ labels all the different reduced density matrices of $|\psi\ra\la\psi|$.
Any $N$-partite pure state that can be written as
$|\psi\ra=|\phi_A\ra\otimes|\phi_{\bar{A}}\ra$ is called
bi-separable, where $A$ denotes a certain subset of ${\mathcal
{H}}_{1}\otimes{\mathcal {H}}_{2}\otimes\cdots\otimes{\mathcal
{H}}_{N}$ and $\bar{A}$ stands for the complement of $A$. States
that are not bi-separable with respect to any bipartition are said
to be genuine multipartite entangled.

For an $N$-partite mixed quantum state,
$\rho=\sum_{i}p_{i}|\psi_{i}\ra\la\psi_{i}| \in {\mathcal
{H}}_{1}\otimes{\mathcal {H}}_{2}\otimes\cdots\otimes{\mathcal
{H}}_{N}$, the corresponding concurrence is given by the convex roof:
\begin{eqnarray}\label{def}
C_{N}(\rho)=\min_{\{p_{i},|\psi_{i}\ra\}}\sum_{i}p_{i}C_{N}(|\psi_{i}\ra\la\psi_{i}|),
\end{eqnarray}
where the minimization runs over all ensembles of pure state
decompositions of $\rho$. A genuine multipartite entangled mixed
state is defined to be one that cannot be written as a convex
combination of biseparable pure states.

{\bf{Theorem}} An N-partite quantum state $\rho \in{\mathcal
{H}}_{1}\otimes{\mathcal {H}}_{2}\otimes\cdots\otimes{\mathcal
{H}}_{N}$ is genuine multipartite entangled if
\begin{widetext}
\begin{eqnarray}\label{theorem}
C_{N}(\rho)>
\left\{\begin{array}{l}
\displaystyle 2^{1-\frac{N}{2}}\sqrt{2^{N}-4+\frac{2}{d}-2\sum_{k=1}^{\frac{N-1}{2}}\frac{\binom{N}{k}}{d^k}},~~~~~~~~~~~~~~ \text{for~odd} ~N,\\[6mm]
\displaystyle
2^{1-\frac{N}{2}}\sqrt{2^{N}-4+\frac{2}{d}-2\sum_{k=1}^{\frac{N}{2}-1}\frac{\binom{N}{k}}{d^k}-\frac{\binom{N}{\frac{N}{2}}}{d^{\frac{N}{2}}}},~~~~\text{for~even}
~N,
\end{array}\right.
\end{eqnarray}
\end{widetext}
where $\binom{N}{k}=N!/(k!(N-k)!)$.

{\sf{Proof}} A general multipartite state $\rho\in {\mathcal
{H}}_{1}\otimes{\mathcal {H}}_{2}\otimes\cdots\otimes{\mathcal
{H}}_{N}$ can be written as \cite{hassan},
\begin{widetext}
\be\nonumber
\ba{rcl}
\rho&=&\displaystyle\frac{1}{d^N}(\left.\otimes_{j=1}^{N}I_{d}
+\sum\limits_{\{\mu_{1}\}}\sum\limits_{\alpha_{1}}
{\mathcal{T}}_{\alpha_{1}}^{\{\mu_{1}\}}\lambda_{\alpha_{1}}^{\{\mu_{1}\}}
+\sum\limits_{\{\mu_{1}\mu_{2}\}}\sum\limits_{\alpha_{1}\alpha_{2}}
{\mathcal{T}}_{\alpha_{1}\alpha_{2}}^{\{\mu_{1}\mu_{2}\}}\lambda_{\alpha_{1}}
^{\{\mu_{1}\}}\lambda_{\alpha_{2}}^{\{\mu_{2}\}}+\sum\limits_{\{\mu_{1}\mu_{2}\mu_{3}\}}\sum\limits_{\alpha_{1}\alpha_{2}\alpha_{3}}
{\mathcal{T}}_{\alpha_{1}\alpha_{2}\alpha_{3}}^{\{\mu_{1}\mu_{2}\mu_{3}\}}\lambda_{\alpha_{1}}
^{\{\mu_{1}\}}\lambda_{\alpha_{2}}^{\{\mu_{2}\}}\lambda_{\alpha_{3}}^{\{\mu_{3}\}}\right.\\[5mm]
&&+\cdots
+\sum\limits_{\{\mu_{1}\mu_{2}\cdots\mu_{M}\}}\sum\limits_{\alpha_{1}\alpha_{2}\cdots\alpha_{M}}
{\mathcal{T}}_{\alpha_{1}\alpha_{2}\cdots\alpha_{M}}^{\{\mu_{1}\mu_{2}\cdots\mu_{M}\}}\lambda_{\alpha_{1}}
^{\{\mu_{1}\}}\lambda_{\alpha_{2}}^{\{\mu_{2}\}}\cdots\lambda_{\alpha_{M}}^{\{\mu_{M}\}}\left.+\cdots
+\sum\limits_{\alpha_{1}\alpha_{2}\cdots\alpha_{N}}
{\mathcal{T}}_{\alpha_{1}\alpha_{2}\cdots\alpha_{N}}^{\{1,2,\cdots,N\}}\lambda_{\alpha_{1}}
^{\{1\}}\lambda_{\alpha_{2}}^{\{2\}}\cdots\lambda_{\alpha_{N}}^{\{N\}}\right.),
\ea \ee
\end{widetext}
where $\lambda_{\alpha_{k}}$ are the $SU(d)$ generators,
$\{\mu_{1}\mu_{2}\cdots\mu_{M}\}$ is a subset of $\{1,2,\cdots,
N\}$, $\lambda_{\alpha_{k}}^{\{\mu_{k}\}}=I_{d}\otimes
I_{d}\otimes\cdots\otimes \lambda_{\alpha_{k}}\otimes
I_{d}\otimes\cdots\otimes I_{d}$ with $\lambda_{\alpha_{k}}$ appearing
at the $\mu_k$th position, $I_d$ is the $d\times d$ identity matrix and
\begin{eqnarray*}
{\mathcal{T}}_{\alpha_{1}\alpha_{2}\cdots\alpha_{M}}
^{\{\mu_{1}\mu_{2}\cdots\mu_{M}\}}=\frac{d^M}{2^{M}}{\rm
Tr}[\rho\lambda_{\alpha_{1}}
^{\{\mu_{1}\}}\lambda_{\alpha_{2}}^{\{\mu_{2}\}}\cdots\lambda_{\alpha_{M}}^{\{\mu_{M}\}}],
\end{eqnarray*}
which can be viewed as the entries of tensors
${\mathcal{T}}^{\{\mu_{1}\mu_{2}\cdots\mu_{M}\}}$.

We start with an N-partite pure quantum state $|\psi\ra$.
Let $||\bullet||$ denote the Euclidean norm for a tensor. After
tedious but straightforward computation, one obtains that for odd
$N$,
\begin{widetext}
\begin{eqnarray*}
\sum_{\alpha=1}^{2^{N}-2}Tr\rho_{\alpha}^2=&2[&\binom{N}{1}\frac{1}{d}+\binom{N-1}{0}\frac{2}{d^2}\sum_{k_1\in\{1,2,\cdots
N\}}||{\mathcal{T}}^{k_1}||^2\\
&+&\binom{N}{2}\frac{1}{d^2}+\binom{N-1}{1}\frac{2}{d^3}\sum||{\mathcal{T}}^{k_1}||^2+\binom{N-2}{0}\frac{2^2}{d^{2*2}}\sum_{k_1k_2}
||{\mathcal{T}}^{k_1k_2}||^2\\
&+&\cdots +\\
&+&\binom{N}{\frac{N-1}{2}}\frac{1}{d^{\frac{N-1}{2}}}+\binom{N-1}{{\frac{N-1}{2}}-1}\frac{2}{d^{\frac{N-1}{2}}+1}\sum||{\mathcal{T}}^{k_1}||^2
+\binom{N-2}{{\frac{N-1}{2}}-2}\frac{2^2}{d^{{\frac{N-1}{2}}+2}}\sum_{k_1k_2}||{\mathcal{T}}^{k_1k_2}||^2\\
&+& \cdots +
\binom{N-{\frac{N-1}{2}}}{0}\frac{2^{\frac{N-1}{2}}}{d^{2*{\frac{N-1}{2}}}}\sum_{k_1\cdots
k_{\frac{N-1}{2}}}||{\mathcal{T}}^{k_1\cdots k_{\frac{N-1}{2}}}||^2
];
\end{eqnarray*}
while for even N,
\begin{eqnarray*}
\sum_{\alpha=1}^{2^{N}-2}Tr\rho_{\alpha}^2=&2[&\binom{N}{1}\frac{1}{d}+\binom{N-1}{0}\frac{2}{d^2}\sum_{k_1\in\{1,2,\cdots
N\}}||{\mathcal{T}}^{k_1}||^2\\
&+&\binom{N}{2}\frac{1}{d^2}+\binom{N-1}{1}\frac{2}{d^3}\sum||{\mathcal{T}}^{k_1}||^2+\binom{N-2}{0}\frac{2^2}{d^{2*2}}\sum_{k_1k_2}||{\mathcal{T}}^{k_1k_2}||^2\\
&+&\cdots\cdots\\
&+&\binom{N}{\frac{N-2}{2}}\frac{1}{d^{\frac{N-2}{2}}}+\binom{N-1}{{\frac{N-2}{2}}-1}\frac{2}{d^{\frac{N-2}{2}}+1}\sum||{\mathcal{T}}^{k_1}||^2
+\binom{N-2}{{\frac{N-2}{2}}-2}\frac{2^2}{d^{{\frac{N-2}{2}}+2}}\sum_{k_1k_2}||{\mathcal{T}}^{k_1k_2}||^2\\
&+&\cdots+\binom{N-{\frac{N-2}{2}}}{0}\frac{2^{\frac{N-2}{2}}}{d^{2*{\frac{N-2}{2}}}}\sum_{k_1\cdots
k_{\frac{N-2}{2}}}||{\mathcal{T}}^{k_1\cdots
k_{\frac{N-2}{2}}}||^2]\\
&+&\binom{N}{\frac{N}{2}}\frac{1}{d^{\frac{N}{2}}}+\binom{N-1}{{\frac{N}{2}}-1}\frac{2}{d^{\frac{N}{2}}+1}\sum||{\mathcal{T}}^{k_1}||^2
+\binom{N-2}{{\frac{N-2}{2}}-2}\frac{2^2}{d^{{\frac{N-2}{2}}+2}}\sum_{k_1k_2}||{\mathcal{T}}^{k_1k_2}||^2\\
&+&\cdots+\binom{N-{\frac{N}{2}}}{0}\frac{2^{\frac{N}{2}}}{d^{2*{\frac{N}{2}}}}\sum_{k_1\cdots
k_{\frac{N}{2}}}||{\mathcal{T}}^{k_1\cdots k_{\frac{N}{2}}}||^2.
\end{eqnarray*}
\end{widetext}

Assume that $|\psi\ra\in{\mathcal {H}}_{1}\otimes{\mathcal
{H}}_{2}\otimes\cdots\otimes{\mathcal {H}}_{N}$ be $R_M|\bar{R}_M$
separable, where $R_M=\{{\mathcal {H}}_{i_1}\otimes {\mathcal
{H}}_{i_2}\otimes\cdots\otimes {\mathcal {H}}_{i_M}\}$ and
$\bar{R}_M$ be the complement of $R_M$. Without loss of generality,
we assume that $1\leq M\leq [\frac{N}{2}]$. As $|\psi\ra$ is a pure
state, we have $Tr(\rho^2_{R_M})=1$, namely,
\begin{widetext}
\begin{eqnarray*}
\frac{2}{d^{M+1}}\sum_{j\in\{1,2,\cdots
M\}}||{\mathcal{T}}^{i_j}||^2+\frac{2^2}{d^{M+2}}\sum_{j,l}||{\mathcal{T}}^{i_ji_l}||^2+\cdots+\frac{2^M}{d^{2M}}||{\mathcal{T}}^{i_1\cdots
i_M}||^2=1-\frac{1}{d^M}.
\end{eqnarray*}

Thus one has that for odd N,
\begin{eqnarray*}
\sum_{\alpha=1}^{2^{N}-2}Tr\rho_{\alpha}^2\geq
2[\binom{N}{1}\frac{1}{d}+\binom{N}{2}\frac{1}{d^2}+\cdots+\binom{N}{\frac{N-1}{2}}\frac{1}{d^{\frac{N-1}{2}}}]+2(1-\frac{1}{d^M}),
\end{eqnarray*}

while for even N,
\begin{eqnarray*}
\sum_{\alpha=1}^{2^{N}-2}Tr\rho_{\alpha}^2\geq
2[\binom{N}{1}\frac{1}{d}+\binom{N}{2}\frac{1}{d^2}+\cdots+\binom{N}{\frac{N-2}{2}}\frac{1}{d^{\frac{N-2}{2}}}]
+\binom{N}{\frac{N}{2}}\frac{1}{d^{\frac{N}{2}}}+2(1-\frac{1}{d^M}).
\end{eqnarray*}

Therefore, we have
\begin{eqnarray*}
C_{N}(|\psi\ra\la\psi|)&=&2^{1-\frac{N}{2}}\sqrt{(2^{N}-2)-\sum_{\alpha}Tr\{\rho_{\alpha}^{2}\}}\\[3mm]
&\leq&\left\{\begin{array}{l}
\displaystyle 2^{1-\frac{N}{2}}\sqrt{2^{N}-4+\frac{2}{d^M}-2\sum_{k=1}^{\frac{N-1}{2}}\binom{N}{k}\frac{1}{d^k}},~~~~~~~~~~~~~~~~~~~~~\text{for~odd} ~N;\\[6mm]
\displaystyle
2^{1-\frac{N}{2}}\sqrt{2^{N}-4+\frac{2}{d^M}-2\sum_{k=1}^{\frac{N}{2}-1}\binom{N}{k}\frac{1}{d^k}-\binom{N}{\frac{N}{2}}\frac{1}{d^{\frac{N}{2}}}},~~~~\text{for~even}
~N.
\end{array}\right.
\end{eqnarray*}

Let $\rho = \sum_i p_i |{\psi_i}\ra \la{\psi_i}|$ be a mixed state
decomposable in terms of an ensemble of biseparable states
$|\psi_i\ra$. By the convexity of the concurrence, $C_N(\rho) \leq
\sum_i p_i C_N(|{\psi_i}\ra\la{\psi_i}|)$. Using the previous
inequality, we get
\begin{eqnarray*}
C_{N}(\rho)&\leq&\left\{\begin{array}{l}
\displaystyle \sum_ip_i2^{1-\frac{N}{2}}\sqrt{2^{N}-4+\frac{2}{d^{M_i}}-2\sum_{k=1}^{\frac{N-1}{2}}\binom{N}{k}\frac{1}{d^k}}, ~~~~~~~~~~~~~~~~~~~~~\text{for~odd} ~N;\\[6mm]
\displaystyle \sum_i
p_i2^{1-\frac{N}{2}}\sqrt{2^{N}-4+\frac{2}{d^{M_i}}-2\sum_{k=1}^{\frac{N}{2}-1}\binom{N}{k}\frac{1}{d^k}-\binom{N}{\frac{N}{2}}\frac{1}{d^{\frac{N}{2}}}},~~~~\text{for~even}
~N;
\end{array}\right.\\[3mm]
&\leq&\left\{\begin{array}{l}
\displaystyle 2^{1-\frac{N}{2}}\sqrt{2^{N}-4+\frac{2}{d}-2\sum_{k=1}^{\frac{N-1}{2}}\binom{N}{k}\frac{1}{d^k}},~~~~~~~~~~~~~~~~~~~~~  \text{for~odd} ~N;\\[6mm]
\displaystyle
2^{1-\frac{N}{2}}\sqrt{2^{N}-4+\frac{2}{d}-2\sum_{k=1}^{\frac{N}{2}-1}\binom{N}{k}\frac{1}{d^k}-\binom{N}{\frac{N}{2}}\frac{1}{d^{\frac{N}{2}}}},~~~~\text{for~even}
~N,
\end{array}\right.
\end{eqnarray*}
\end{widetext}
where in the last inequality we have used the fact that
$\frac{2}{d}\geq \frac{2}{d^M}$ for any $M\geq 1$. \hfill \rule{1ex}{1ex}

{\sf{Remark 1:}} The lower bound for the multipartite concurrence
presented in the above theorem together with the fact that
$C_{N}(\rho)=0$ for fully separable states supply a kind of classification
for multipartite entanglement which only depends the dimensions and the
number of subsystems.

The lower bound (\ref{theorem}) of the multipartite concurrence
$C_{N}(\rho)$ of a state $\rho$ presents a sufficient condition for
a state to be genuine multipartite entangled. Besides, if we take
$N=3$ and $d=2$ and consider any bi-separable pure state
$|\psi_{123}\ra=|\psi_{12}\ra\otimes|\psi_3\ra$ with
$|\psi_{12}\ra=(|00\ra+|11\ra)/\sqrt{2}$, then the concurrence
$C_3(|\psi_{123}\ra)$ is 1, which is just the maximal value of the
bound (\ref{theorem}) for any bi-separable states. Thus this bound
is tight in this case.

The theorem gives an effective way to detect genuine multipartite
entanglement by estimating the multipartite concurrence of a state.
Generally, it is difficult to calculate analytically the
multipartite concurrence of a given state. Nevertheless, there have
been many results on the lower bounds of the multipartite
concurrence for mixed states \cite{jpali,gao,zhao,zhang}. From our
theorem these bounds give rise to criteria of the genuine
multipartite entanglement. To employ our theorem for detailed
applications, we first present a new lower bound of multipartite
concurrence in the following.

For a pure N-partite quantum state $|\psi\ra\in {\mathcal
{H}}_{1}\otimes{\mathcal {H}}_{2}\otimes\cdots\otimes{\mathcal
{H}}_{N}$, the bipartite concurrence with respect to the bipartite decomposition
$\alpha|\bar{\alpha}$ is defined by
\begin{eqnarray}\label{xx}
C_{2}^{\alpha}(|\psi\ra\la\psi|)=\sqrt{2(1-Tr\{\rho_{\alpha}^{2}\})},
\end{eqnarray}
where $\rho_{\alpha}=Tr_{\bar{\alpha}}\{|\psi\ra\la\psi|\}$ is the reduced
density matrix of $\rho=|\psi\ra\la\psi|$ by tracing over the subsystem $\bar{\alpha}$.
For a mixed multipartite quantum state,
$\rho=\sum_{i}p_{i}|\psi_{i}\ra\la\psi_{i}|$,
the corresponding bipartite concurrence is given by
\begin{eqnarray}\label{def1}
C_{2}^{\alpha}(\rho)=\min_{\{p_{i},|\psi_{i}\}\ra}\sum_{i}p_{i}C_{2}^{\alpha}(|\psi_{i}\ra\la\psi_{i}|),
\end{eqnarray}
where the minimization runs over all ensembles of pure state decompositions of $\rho$.
We have the following results:

{\bf{Proposition}} For any mixed multipartite quantum state $\rho
\in {\mathcal {H}}_{1}\otimes{\mathcal{H}}_{2}\otimes\cdots\otimes{\mathcal {H}}_{N}$,
the multipartite concurrence (\ref{def}) is bounded by
\begin{eqnarray}\label{Prop}
C_{N}(\rho)\geq
2^{\frac{1-N}{2}}\sqrt{\sum_{\alpha=1}^{2^N-2}(C_2^{\alpha}(\rho))^2}.
\end{eqnarray}

{\sf{Proof}} We start the proof with a pure state $|\psi\ra\in
{\mathcal {H}}_{1}\otimes{\mathcal
{H}}_{2}\otimes\cdots\otimes{\mathcal {H}}_{N}$. According to the
definition of the multipartite concurrence, one obtains that
\begin{eqnarray*}\label{proof1}
C_{N}(|\psi\ra\la\psi|)&=&2^{1-\frac{N}{2}}\sqrt{(2^{N}-2)-\sum_{\alpha=1}^{2^N-2}Tr\{\rho_{\alpha}^{2}\}}\nonumber\\
&=&2^{1-\frac{N}{2}}\sqrt{\sum_{\alpha=1}^{2^N-2}x_{\alpha}^2},
\end{eqnarray*}
where we have set $x_{\alpha}=\sqrt{1-Tr\{\rho_{\alpha}^{2}\}}$.

For any mixed state $\rho\in {\mathcal {H}}_{1}\otimes{\mathcal
{H}}_{2}\otimes\cdots\otimes{\mathcal {H}}_{N}$,
assume that $\{p_i, |\psi_i\ra\}$ is the optimal ensemble of pure state decomposition
such that $C_{N}(\rho)=\sum_i p_i C_N(|\psi_i\ra\la\psi_i|)$.
Using the Minkowski inequality, one derives that
\begin{eqnarray*}
C_{N}(\rho)&=&\sum_ip_iC_N(|\psi_i\ra\la\psi_i|)=2^{1-\frac{N}{2}}\sum_ip_i\sqrt{\sum_{\alpha=1}^{2^N-2}x^2_{i\alpha}}\\
&\geq&2^{1-\frac{N}{2}}\sqrt{\sum_{\alpha=1}^{2^N-2}(\sum_ip_ix_{i\alpha})^2}\\
&\geq&2^{\frac{1-N}{2}}\sqrt{\sum_{\alpha=1}^{2^N-2}(C_2^{\alpha}(\rho))^2},
\end{eqnarray*}
which proves the proposition.\hfill \rule{1ex}{1ex}

{\sf{Remark 2:}} (\ref{Prop}) is a kind of monogamy
inequality \cite{monogamy} for multipartite entanglement in terms of the
difference between total entanglement and the bipartite entanglement.
 Let $\rho\in {\mathcal {H}}_{A}\otimes{\mathcal
{H}}_{B}\otimes{\mathcal {H}}_{C}$ for instance. (\ref{Prop}) is
then represented by
$(C_3(\rho))^2\geq(C_{2}^A(\rho))^2+(C_{2}^B(\rho))^2+(C_{2}^C(\rho))^2$.
Combining the monogamy inequality derived in \cite{monogamy} one has
$(C_3(\rho))^2\geq
(C_{2}(\rho_{AB}))^2+(C_{2}(\rho_{AC}))^2+(C_{2}(\rho_{BC}))^2$,
where $\rho_{AB},\rho_{AC}$ and $\rho_{BC}$ are the reduced matrices
of $\rho$.

In (\ref{Prop}) the lower bound of multipartite concurrence
$C_{N}(\rho)$ is given by the concurrences of bipartite
partitions. These bipartite concurrences can be estimated in many
operational approaches \cite{chenk,zhao,zhang}.

By using our theorem and proposition we now investigate the genuine multipartite entanglement
by detailed examples. Let us consider the state in three qutrits systems,
$$
\rho_{GGHZ}=\frac{x}{27}I_{27}+(1-x)|GGHZ\ra\la GGHZ|,
$$
where $|GGHZ\ra$ is a generalized GHZ state,
$|GGHZ\ra=(|000\ra+|111\ra+|222\ra)/\sqrt{3}$.
By using the lower bounds for bipartite states \cite{chenk}, it is direct to
obtain the lower bound (\ref{Prop}) of the concurrence $C_3(\rho)$. The genuine multipartite
entanglement is then detected for $0<x<0.16515$, which is better than
the result from the theorem 1 in \cite{vicente3}: $0<x<0.10557$ (or that in
\cite{claude} as they are of the same power for detecting genuine multipartite
entanglement of $\rho_{GGHZ}$).

As another example we consider
$\rho_{GHZ}=\frac{x}{8}I_{8}+(1-x)|GHZ\ra\la GHZ|$, where
$|GHZ\ra=(|000\ra+|111\ra)/\sqrt{2}$. The lower bound in
(\ref{theorem}) is given by $\sqrt{2-\frac{2}{d}}=1$. By using the
lower bound for bipartite concurrence in \cite{zhao}, we have
$C_3(\rho)\geq \sqrt{\frac{3}{2}}\frac{1-3x}{1+3x}$. From our
theorem the genuine multipartite entanglement is detected for
$x<0.033$. If we employ the lower bound of concurrence in
\cite{jpali}, genuine multipartite entanglement is detected by our
theorem for $x<0.1468$, which is also better than the range $x<0.13$
obtained by using the theorem 1 in \cite{vicente3}. One may always
enhance the power of detecting genuine multipartite entanglement by
employing better lower bounds of multipartite concurrence. Here for
the state $\rho_{GHZ}$, its lower bound of concurrence from
\cite{gao} is given by $C_3(\rho_{GHZ})\geq
-\frac{1}{2}+\frac{3-3x}{4}+\frac{2-2x+x^2}{4\sqrt{2}}$. Therefore
for $0\leq x\leq 0.190211$, the lower bound from our proposition is
better than that from \cite{gao} (see Fig. 1).

\begin{figure}[h]
\begin{center}
\resizebox{9cm}{!}{\includegraphics{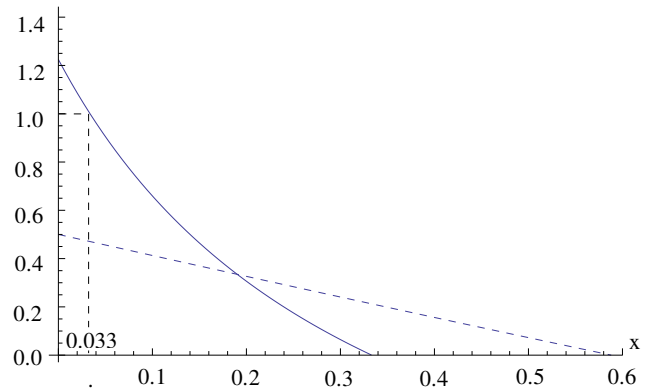}}
\end{center}
\caption{Lower bound of concurrence from proposition (solid line)
and that from \cite{gao} (dashed line). The solid line shows that
for $x<0.033$, $C_3(\rho_{GHZ})>1$ and the state is genuine
multipartite entangled. For $0\leq x\leq 0.190211$, the lower bound
from our proposition supplies a better estimation of concurrence
than that from \cite{gao}. \label{fig1}}
\end{figure}

Let us further consider the Dur-Cirac-Tarrach state \cite{dct},
$$
\rho_{DCT}=\sum_{\sigma=\pm}\lambda^{\sigma}_0\vert\psi^{\sigma}_0\rangle\langle\psi^{\sigma}_0\vert
+\sum_{k=1}^3\lambda_k(\vert\psi^{+}_k\rangle\langle\psi^{+}_k\vert+\vert\psi^{-}_k\rangle\langle\psi^{-}_k\vert),
$$
where
$\vert\psi^{\pm}_0\rangle=\frac{1}{\sqrt{2}}(|000\ra\pm|111\ra),
\vert\psi^{\pm}_j\rangle=\frac{1}{\sqrt{2}}(\vert
j\ra_{AB}|0\rangle_C\pm\vert (3-j)\ra_{AB}|0\rangle_C)$,
$|j\ra_{AB}=|j_1\ra_A|j_2\ra_B$ with $j=j_1j_2$ in binary notation.
Take $\lambda^+_0=\frac{1}{6}$, $\lambda^-_0=\frac{1}{2}$,
$\lambda_{1}=\lambda_{2}=\lambda_{3}=\frac{1}{18}$. From Ref.
\cite{gao} the lower bound of concurrence is given by $C(\rho_{DCT}
)\geq 0.3143$, where the difference of a constant factor $\sqrt{2}$
in defining the concurrence for pure states has already been taken
into account. From our proposition and using the lower bound for
bipartite concurrence in \cite{zhao}, we obtain $C(\rho_{DCT} )\geq
0.3499$. Therefore, the lower bound presented in the proposition is
better than the lower bound in Refs. \cite{gao} in detecting the
full separability of the three-qubit mixed state $\rho_{DCT}$.

In summary, for tripartite quantum systems, a state $\rho
\in{\mathcal {H}}_{1}\otimes{\mathcal {H}}_{2}\otimes{\mathcal
{H}}_{3}$ is genuine multipartite entangled if $C_{3}(\rho)>
\sqrt{2-\frac{2}{d}}$. We have the relationship between the property
of entanglement and the value of concurrence, see Fig. 2. Here we
show the detailed processes of detecting genuine entanglement for
arbitrary quantum states $\rho\in{\mathcal {H}}_{1}\otimes{\mathcal
{H}}_{2}\otimes{\mathcal {H}}_{3}$ by using the theorem and the
proposition. One can detect the genuine multipartite entanglement for quantum states
in any $N$ partite systems with arbitrary dimensions.

Step 1: Treat $\rho\in{\mathcal {H}}_{1}\otimes{\mathcal
{H}}_{2}\otimes{\mathcal {H}}_{3}$ in terms of bipartite cuts:
${\mathcal {H}}_{1}|{\mathcal {H}}_{2}{\mathcal {H}}_{3}$,
${\mathcal {H}}_{2}|{\mathcal {H}}_{1}{\mathcal {H}}_{3}$ and
${\mathcal {H}}_{3}|{\mathcal {H}}_{1}{\mathcal {H}}_{2}$. Compute
the lower bound of concurrence for ``bipartite" quantum state
$\rho$. For the example $\rho_{GGHZ}$, we have selected the lower
bound for concurrence given in \cite{chenk}. Indeed, any valid lower bound
for bipartite concurrence (such as that in \cite{vic,ger,ou,zhao})
is adoptable to detect genuine multipartite entanglement.

Step 2: By the proposition, one can compute the lower bound of
concurrence $C_3(\rho)$ (denoted as $LC_3(\rho)$) by summing all the
squared lower bounds of ``bipartite" concurrence and then taking a
square root.

Step 3: Compare $LC_3(\rho)$ derived in the above step (or that has
been derived directly from the lower bound of $C_3(\rho)$ such as
that in \cite{multicon1,multicon2,zhu}) with the lower bound in the
theorem for $N=3$, i.e. $\sqrt{2-\frac{2}{d}}$. If
$LC_3(\rho)>\sqrt{2-\frac{2}{d}}$, genuine multipartite
entanglement is detected.

\begin{figure}[h]
\begin{center}
\resizebox{7cm}{!}{\includegraphics{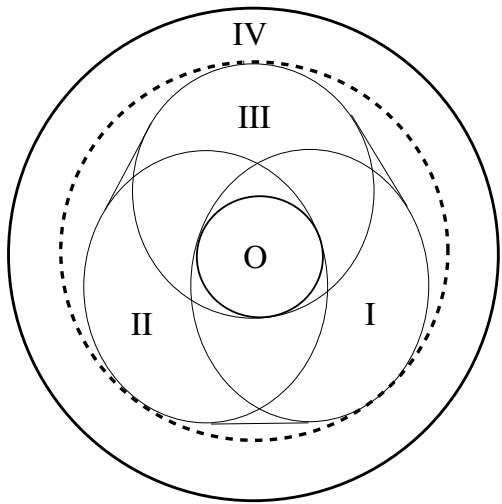}}
\end{center}
\caption{Quantum states located in the area between the largest
circle and the dashed circle are genuine multipartite entangled with
concurrence $C(\rho)>\sqrt{2-\frac{2}{d}}$, part IV. Bipartite
separable states, $0<C(\rho)\leq\sqrt{2-\frac{2}{d}}$, in parts I,
II, III and their linear superposition. And $C(\rho)=0$, fully
separable states in part $0$. The rest undetected part is genuine
multipartite entangled with $0<C(\rho)\leq\sqrt{2-\frac{2}{d}}$.\label{fig2}}
\end{figure}

To detect the genuine multipartite entanglement and measure the
multipartite entanglement are basic and fundamental problems in
quantum information science. We have investigated the relations
between genuine multipartite entanglement and the multipartite
concurrence. It has been shown that if the multipartite concurrence
is larger than a constant depending only on the dimensions and the
number of the subsystems, the state must be genuine multipartite
entangled. We have also derived an analytical and effective lower
bound of multipartite concurrence, which contributes not only to the
detection of genuine multipartite entanglement, but also to the
estimation of multipartite entanglement. In \cite{ksep}, the quantum
$k$-separability for multipartite quantum systems have been studied.
Our method can be also applied to this issue. Besides, the detection
of genuine multipartite entanglement for continuous variable systems
\cite{gcv} may be similarly investigated by bounding the
multipartite concurrence.

\bigskip
\noindent{\bf Acknowledgments}\, \, This work is supported by the
NSFC 11105226, 11275131; the Fundamental Research Funds for the
Central Universities No. 15CX08011A, No. 24720122013 and the
Project-sponsored by SRF for ROCS, SEM.


\smallskip
\begin{thebibliography}{99}
\bibitem{nielsen}M.A. Nielsen and I.L. Chuang, Quantum Computation and Quantum
Information. Cambridge: Cambridge University Press, (2000).

\bibitem{di} See, for example, D.P. Di Vincenzo, Science
270,255(1995).

\bibitem{dct} W. D$\ddot{u}$r, J. I. Cirac, and R. Tarrach, Phys. Rev. Lett. 83,
3562 (1999).

\bibitem{huber} M. Huber and R. Sengupta, Phys. Rev. Lett. 113, 100501
(2014).

\bibitem{vicente3} J.I. de Vicente, M. Huber, Phys. Rev. A, 84, 062306
(2011).

\bibitem{mule1} R. Horodecki et al., Rev. Mod. Phys. 81, 865 (2009).

\bibitem{mule2} H.J. Briegel et al., Nat. Phys. 5, 19 (2009).

\bibitem{mule3} A. Sen(De) and U. Sen, Phys. News 40, 17 (2010),
arXiv:1105.2412.

\bibitem{mule4} M. $\dot{Z}$ukowski et al., Acta Phys. Pol. 93, 187 (1998); M. Hillery
et al., Phys. Rev. A 59, 1829 (1999); R. Demkowicz-Dobrzanski et
al., ibid. 80, 012311 (2009); N. Gisin et al., Rev. Mod. Phy. 74,
145 (2002).

\bibitem{mule5} R. Cleve et al., Phys. Rev. Lett. 83, 648 (1999); A. Karlsson et
al., Phys. Rev. A 59, 162 (1999).


\bibitem{12} M. Huber, F. Mintert, A. Gabriel, and
B. C. Hiesmayr, Phys. Rev. Lett. 104, 210501 (2010).



\bibitem{wu} J.Y. Wu, H. Kampermann, D. Bru${\ss}$, C. Klockl, and M. Huber,
Phys. Rev. A 86, 022319 (2012).

\bibitem{huber1} M. Huber, M. Perarnau-Llobet, J.I.
de Vicente, Phys. Rev. A 88, 042328 (2013).

\bibitem{sperling} J. Sperling, W. Vogel, Phys. Rev. Lett. 111, 110503 (2013).



\bibitem{14} C. Eltschka and J. Siewert, Phys. Rev. Lett. 108, 020502
(2012).

\bibitem{15} B. Jungnitsch, T. Moroder, and O. G$\ddot{u}$hne, Phys. Rev. Lett. 106,
190502 (2011).



\bibitem{claude} C. Klckl, M. Huber, Phys. Rev. A 91, 042339
(2015).

\bibitem{horo} M. Markiewicz, W. Laskowski, T. Paterek, and M. $\dot{Z}$ukowski
Phys. Rev. A 87, 034301 (2013).

\bibitem{ma1} Z.H. Ma, Z.H. Chen, J.L. Chen, C.
Spengler, A. Gabriel, and M. Huber, Phys. Rev. A 83, 062325(2011).

\bibitem{ma2} Z.H. Chen, Z.H. Ma, J.L. Chen, and S. Severini, Phys. Rev. A 85, 062320
(2012).

\bibitem{gaot1} Y. Hong, T. Gao, and F.L. Yan, Phys. Rev. A 86, 062323 (2012).

\bibitem{gaot2} T. Gao,. F.L. Yan, and S.J. van Enk, Phys. Rev. Lett. 112, 180501
(2014).

\bibitem{bellgme} J.D. Bancal, N. Gisin, Y.C. Liang, and S. Pironio, Phys. Rev.
Lett. 106, 250404 (2011).

\bibitem{wootters} W.K. Wootters, Phys. Rev. Lett. 80, 2245(1998).

\bibitem{multicon1} L. Aolita and F. Mintert, Phys. Rev. Lett. 97,
050501(2006).

\bibitem{multicon2} A.R.R. Carvalho, F. Mintert, and A. Buchleitner, Phys. Rev. Lett.
93, 230501(2004).

\bibitem{chenk} K. Chen, S. Albeverio and S.M. Fei, Phys. Rev. Lett. 95,
040504(2005).

\bibitem{vicente1} J.I. de Vicente, Quantum Inf. Comput. 7, 624(2007).


\bibitem{vicente2} J.I. de Vicente, J. Phys. A: Math. and Theor., 41, 065309(2008).


\bibitem{hassan} A.S. M. Hassan, P. S. Joag, Quant. Inf. Comput. 8, 0773(2008).

\bibitem{ming} M. Li, J. Wang, S.M. Fei and X.Q. Li-Jost, Phys. Rev. A, 89,022325(2014).

\bibitem{hassan1} A.S. M. Hassan, P. S. Joag, Phys. Rev. A, 77, 062334
(2008).


\bibitem{hassan2} A.S. M. Hassan, P. S. Joag, Phys. Rev. A, 80, 042302
(2009).


\bibitem{mingbell} M. Li and S.M. Fei, Phys. Rev. A, 86, 052119
(2012).

\bibitem{jpali} M. Li, S.M. Fei  and Z.X. Wang, J. Phys. A, Math. Theor.£¬42, 145303(2009).


\bibitem{gao} X.H. Gao, S.M. Fei, and K. Wu, Phys. Rev. A 74, 050303 (2006).



\bibitem{zhao} M.J. Zhao, X.N. Zhu, S.M. Fei, and X.Q. Li-Jost,
Phys. Rev. A 84, 062322 (2011).

\bibitem{zhang} C.J. Zhang, Y.S. Zhang, S. Zhang and G.C. Guo, Phys. Rev. A
76, 012334(2007).


\bibitem{monogamy} V. Coffman, J. Kundu, and W. K. Wootters, Phys. Rev. A
61, 052306 (2000); T.J. Osborne and F. Verstraete, Phys. Rev. Lett.
96, 220503 (2006); B. Regula, S.D. Martino, S. Lee, and G. Adesso,
Phys. Rev. Lett. 113, 110501(2014).

\bibitem{vic} J. I. de Vicente, Phys. Rev. A 75, 052320(2007).

\bibitem{ger} E. Gerjuoy, Phys. Rev. A 67,052308(2003).

\bibitem{ou} Y. C. Ou, H. Fan, and S. M. Fei, Phys. Rev. A 78, 012311(2008).

\bibitem{zhu} X. N. Zhu, M. J. Zhao and S. M. Fei, Phys. Rev. A 86,
022307(2012).

\bibitem{ksep} A. Gabriel, B. C. Hiesmayr, M. Huber, Quant. Inf. Comput. 10,
829(2010).

\bibitem{gcv} G. Adesso and F. Illuminati, J. Phys. A: Math. Theor. 40,
7821(2007).

\end{thebibliography}
\end{document}